\documentclass[aps,prd,twocolumn,superscriptaddress,nofootinbib]{revtex4-1}

\usepackage{latexsym}
\usepackage{amsmath}
\usepackage{amssymb}
\usepackage{amsfonts}
\usepackage{bm}
\usepackage{physics}

\usepackage{color}
\definecolor{purple}{rgb}{0.5,0,0.5}
\definecolor{blue}{rgb}{0.0,0,0.9}
\definecolor{prdblue}{rgb}{0.133,0.118,0.498}
\usepackage[colorlinks=true, pdfstartview=FitV, linkcolor=prdblue, citecolor= prdblue, urlcolor=prdblue]{hyperref}

\usepackage{supertabular} 
\usepackage{placeins}
\usepackage{epsfig}
\usepackage{graphicx}

\usepackage{soul} 
\usepackage{color}

\begin{document}

\title{Revisiting the proton-antiproton scattering using a constituent-quark-model based coupled-channels calculation}

\author{P. G. Ortega}
\email[]{pgortega@usal.es}
\affiliation{Departamento de F\'isica Fundamental, Universidad de Salamanca, E-37008 Salamanca, Spain}
\affiliation{Instituto Universitario de F\'isica 
Fundamental y Matem\'aticas (IUFFyM), Universidad de Salamanca, E-37008 Salamanca, Spain}

\author{D. R. Entem}
\email[]{entem@usal.es}
\affiliation{Departamento de F\'isica Fundamental, Universidad de Salamanca, E-37008 Salamanca, Spain}
\affiliation{Instituto Universitario de F\'isica
Fundamental y Matem\'aticas (IUFFyM), Universidad de Salamanca, E-37008 Salamanca, Spain}

\author{F. Fern\'andez}
\email[]{fdz@usal.es}
\affiliation{Instituto Universitario de F\'isica 
Fundamental y Matem\'aticas (IUFFyM), Universidad de Salamanca, E-37008 Salamanca, Spain}

\author{J. Segovia}
\email[]{jsegovia@upo.es}
\affiliation{Departamento de Sistemas F\'isicos, Qu\'imicos y Naturales, Universidad Pablo de Olavide, E-41013 Sevilla, Spain}

\date{\today}

\begin{abstract}
Motivated by the last experimental and theoretical advances in the analysis of possible baryonium resonances, the $X(1835)$ and their partners, we perform a constituent-quark-model based coupled-channels calculation of the proton-antiproton scattering in order to analyze the possible existence of bound states or near-threshold structures.
The used $N\overline N$ potential is derived from a $G$-parity transformation of the quark-model-based $NN$ interaction which has described well deuteron properties, $NN$ phase shifts, and even hadron-hadron phenomenology. The additional $N\overline N$ annihilation is taken into account by a complex phenomenological potential whose real part is generated by one-pion and one-gluon exchange annihilation potentials and its imaginary part is an energy-independent potential of Gaussian form. Then, all the parameters of the interaction are constrained by the $NN$ sector except those determining the imaginary part of the annihilation potential.
Our study concludes that the nucleon-antinucleon dynamics is complex and rich in scattering singularities near the proton-antiproton and neutron-antineutron thresholds.
\end{abstract}


\maketitle


\section{INTRODUCTION}
\label{sec:intro}

The BESII collaboration observed at the beginning of the millennium a hint for a resonance, called $X(1835)$, in the $J/\psi\to \gamma p\bar{p}$~\cite{BES:2003aic} and $J/\psi\to \pi^+\pi^-\eta^\prime$~\cite{BES:2005ega} decays. The two reported observations of the $X(1835)$ bump stimulated more experimental efforts because both provide similar values for the mass, around $1840\,\text{MeV}$, but quite different results for the decay width: Ref.~\cite{BES:2003aic} observed a very narrow resonance of about some MeV whereas Ref.~\cite{BES:2005ega} determined a width in the order of a few tens of MeV. Thereafter, CLEO~\cite{CLEO:2010fre} and BESIII~\cite{BESIII:2010vwa, BESIII:2010gmv} collaborations confirmed the existence of the $X(1835)$ bump and, moreover, both found that it has the quantum numbers $J^{PC}=0^{-+}$~\cite{BESIII:2011aa, BESIII:2015xco}. In the mean time, a new decay mode $3(\pi^+\pi^-)$ of $X(1835)$ was observed by the BESIII collaboration in the decay $J/\psi\to \gamma 3(\pi^+\pi^-)$~\cite{BESIII:2013sbm}; the mass and width were measured to be $1842.2\pm4.2_{-2.6}^{+7.1}\,\text{MeV}$ and $83\pm14\pm11\,\text{MeV}$, respectively. This is to say, while the mass is again compatible with former measurements, the measured width is in better agreement with the first observations than the ones reported latter, \emph{e.g.} Ref.~\cite{BESIII:2010gmv} measured $190.1\pm9.0\,\text{MeV}$.

A more recent analysis of $J/\psi\to \gamma\pi^+\pi^-\eta^\prime$ performed by the BESIII collaboration~\cite{BESIII:2016fbr} observed a marked change in the line-shape of $X(1835)\to \pi^+\pi^-\eta^\prime$ at the $p\bar{p}$ threshold, which could be originated from either the opening of an additional $p\bar{p}$ decay channel (threshold effect) or the interference between two different resonance contributions. In order to take into account both possibilities, they fitted the line-shape following a couple of models. In the first one, experimentalists considered just one structure deformed by the opening of $p\bar p$ threshold and obtained
\begin{subequations}
\begin{align}
M_{\rm pole} &= 1909.5\pm 15.9^{+9.4}_{-27.5}\,\text{MeV} \,, \\
\Gamma_{\rm pole} &= 273.5\pm 21.4^{+6.1}_{-64.0}\,\text{MeV} \,.
\end{align}
\end{subequations}
In the second case, they considered two structures that interfere near the $p\bar p$ threshold. Such resonances were named $X(1835)$ and $X(1870)$, and they have the following parameters
\begin{subequations}
\begin{align}
M_{\rm pole} &= 1825.3\pm 2.4^{+17.3}_{-2.4}\,\text{MeV} \,, \\
\Gamma_{\rm pole} &= 245.2\pm 13.1^{+4.6}_{-9.6}\,\text{MeV} \,,
\end{align}
\end{subequations}
for the $X(1835)$, and
\begin{subequations}
\begin{align}
M_{\rm pole} &= 1870.2\pm 2.2^{+2.3}_{-0.7}\,\text{MeV}\,, \\
\Gamma_{\rm pole} &= 13.0\pm 6.1^{+2.1}_{-3.8}\,\text{MeV} \,.
\end{align}
\end{subequations}
for the $X(1870)$.

In Autumn 2023, the BESIII collaboration published an article~\cite{BESIII:2023vvr} in which a similar phenomenon to that found in $J/\psi\to \gamma\pi^+\pi^-\eta^\prime$ decay exists around the $p\bar{p}$ threshold in the invariant mass spectrum of $3(\pi^+\pi^-)$ when $J/\psi$ decays into $\gamma 3(\pi^+\pi^-)$. The size of the sample was about fifty times greater than that used in Ref.~\cite{BESIII:2013sbm}, and they found the named $X(1840)$ and $X(1880)$ states with the following parameters:
\begin{subequations}
\begin{align}
M_{\rm pole} &= 1832.5\pm 3.1\pm 2.5\,\text{MeV} \,,\\
\Gamma_{\rm pole} &= 80.7\pm 5.2\pm 7.7\,\text{MeV} \,,
\end{align}
\end{subequations}
and
\begin{subequations}
\begin{align}
M_{\rm pole} &= 1882.1\pm 1.7\pm 0.7\,\text{MeV} \,,\\
\Gamma_{\rm pole} &= 30.7\pm 5.5\pm 2.4\,\text{MeV} \,.
\end{align}
\end{subequations}

Finally, it is worth noting that an enhancement at $X(1810)$ has also been observed in the reaction $J/\psi \to \gamma \omega \phi$~\cite{BESIII:2012rtd}. The quantum numbers of this state have been identified as $J^{PC} = 0^{++}$, and although there is some suggestion that it may be connected to previously observed states, this relationship remains uncertain.

On the theoretical side, similar efforts to interpret the experimental signals have been develop in the last $20$ years. Their simplest interpretation is that they are baryonium bound states~\cite{Datta:2003iy, Chang:2004us, Ding:2005ew, Loiseau:2005cv, Wang:2006sna, Dedonder:2009bk}; however, the experimental data has been analyzed with other different point of views such as proton-antiproton final state interactions~\cite{Sibirtsev:2004id, Ding:2005ew, Liu:2009vm}, baryonium states with large gluon content~\cite{Ding:2005gh}, pseudoscalar glueballs~\cite{Li:2005vd, Kochelev:2005vd, Hao:2005hu} or even radial excitations of the $\eta^\prime$ meson~\cite{Huang:2005bc}. 

Motivated by the last experimental and theoretical advances in the analysis of the $X(1835)$ signal, and their derivatives, we revisit and update the perturbative calculation shown in Ref.~\cite{Entem:2006dt} where an investigation of the possible existence of bound states or near-threshold resonances in proton-antiproton scattering, as well as energy shifts in protonium, was performed. The used $N\overline N$ potential is derived from a $G$-parity transformation of the quark-model-based $NN$ interaction reported in Ref.~\cite{Entem:2000mq} and successfully applied to deuteron properties, $NN$ phase shifts, and hadron-hadron phenomenology~\cite{Entem:2000mq, Valcarce:2005em}. The additional $N\overline N$ annihilation contribution is taken into account as in Ref.~\cite{Entem:2006dt}  where it is given by a complex phenomenological potential whose real part is generated by one-pion and one-gluon exchange annihilations and its imaginary part is an energy-independent potential of Gaussian form. In this way, all the parameters of the interaction are constrained by the $NN$ sector except the two describing the imaginary potential. The novelty of the calculation presented here with respect to that of Ref.~\cite{Entem:2006dt} lies in part in the fact that this calculation is non-perturbative, as will be seen later.

It is worth mentioning here that the real part of the $N\overline N$ interaction is usually derived as a $G$-parity transformation of the $NN$ one (see Ref.~\cite{Klempt:2002ap} and references therein). However, this might not be straightforwardly applied to the short-range interaction because it is described phenomenologically~\cite{Klempt:2002ap}, ignoring quark degrees of freedom, which should be relevant at this scale. In quark-based $NN$ interactions, the short-range repulsion comes from the antisymmetry between quarks. This fact has important consequences on the $N\overline N$ potentials as compared with those derived from one-boson exchange models. In the meson-exchange picture, the central force is basically provided by the $\sigma$- and $\omega$-exchange interactions. They have opposite sign for the $NN$ system but add coherently in the $N\overline N$ one. Moreover, the spin-orbit force coming from the $\sigma$- and $\omega$-exchange interactions adds in $NN$ and cancels in $N\bar N$. The $\omega$-exchange contribution is replaced in quark-based models by the antisymmetry effect, which is not present in $N\overline N$. Therefore, quark-based $N\overline N$ interaction may look different from the conventional one described by one-boson exchange potentials. Furthermore, the $NN$ one-pion-exchange tensor interaction is attenuated also by the antisymetry operation, and not by $\rho$-exchanges.

Note also that the imaginary part of the $N\overline{N}$ interaction presents its own theoretical challenges. An annihilation potential is required, which has traditionally been modeled using baryon exchanges~\cite{Klempt:2002ap}, though the most accurate results have been achieved through phenomenological approaches. Alternatively, a quark-level description of the $N\overline{N}$ annihilation can be employed, describing it in terms of quark rearrangements driven by quark-quark interactions. However, certain parameters still need to be refined, and in this regard, the new experimental data from BESIII offers an opportunity to update our previous studies on the $N\overline{N}$ interaction and the potential existence of proton-antiproton bound states.

The manuscript is then organized as follows. After this introduction, the details of the constituent quark model and the theoretical framework are presented in section~\ref{sec:theory}. The analysis and discussion of the results is included in section~\ref{sec:results}. Finally, we summarize and provide some conclusions in Sec.~\ref{sec:summary}.


\section{THEORETICAL FRAMEWORK}
\label{sec:theory}

We use a constituent quark model (CQM) inspired in the phenomenology of QCD at low and medium energy range~\cite{Vijande:2004he, Segovia:2013wma, Fernandez:2019ses}. Its core feature emerges from the spontaneous breaking of the original $SU(3)_L\otimes SU(3)_R$ symmetry of the QCD Lagrangian at some momentum scale due to the small but non-zero current quark masses. This dynamically generates a constituent quark mass, which is estimated to be around $300\,\text{MeV}$~\cite{Diakonov1996}; moreover, it provides a natural coupling for quarks and pions due to Goldstone theorem. An effective Lagrangian, invariant under chiral rotations, can be built as~\cite{Diakonov1996}
\begin{equation}\label{lagrangian}
{\mathcal L}
=\overline{\psi }(i\, {\slash\!\!\! \partial} -M(q^{2})U^{\gamma_{5}})\,\psi \,,
\end{equation}
being $\psi$ the quark spinor, $U^{\gamma_5} = \exp(i f_\pi^{-1} \gamma_{5} \lambda_{a} \phi ^{a})$ the matrix of Goldstone boson fields, with $\phi^a=\{\vec \pi,K_i,\eta_8\}$ the pseudoscalar fields ($i=1,\ldots,4$), and $M(q^2)$ the dynamical (constituent) quark mass which is frozen at low momenta and vanishes at large ones. The momentum dependence of the constituent quark mass can be parametrized as $M(q^2)=m_q F(q^2)$, where $m_{q}\simeq300\,\text{MeV}$, and
\begin{equation}
F(q^{2})=\left[ \frac{\Lambda_{\chi SB}^{2}}{\Lambda_{\chi SB}^{2}+q^{2}} \right]^{{\frac12} } \,.
\end{equation}
The cut-off $\Lambda_{\chi SB}$ is a parameter of the model that fixes the chiral symmetry breaking scale. 

As we already mentioned, the Goldstone theorem provides a natural way of coupling the quarks and pions. Indeed, when the Goldstone boson matrix field $U^{\gamma_5}$ is expanded,
\begin{equation}
U^{\gamma _{5}} = 1 + \frac{i}{f_{\pi }} \gamma^{5} \lambda^{a} \phi^a - \frac{1}{2 f_{\pi }^{2}} \phi^a \phi^a + \ldots \,,
\end{equation}
each term provides an ingredient of the quark model. The first term describes the constituent quark mass, the second term is the one-pion exhange interaction between quarks and the third term represents a two-pion exchange diagram which can be modeled as a scalar $\sigma$-exchange interaction~\cite{Valcarce:1995dm}. From here, it is straightforward to obtain the non-relativistic potentials in the static approximation for the $N\overline N$ system as
\begin{align}
V^{PS}_{ij} (\vec{q}\,) &= \frac{1}{(2\pi)^3} \,\frac{g_{ch}^2}{4m_q^2}
\,\frac{\Lambda^2_{\chi SB}}{\Lambda^2_{\chi SB}+q^2} \nonumber \\
&
\times \frac{(\vec{\sigma}_i \cdot \vec{q}\,)(\vec{\sigma}_j \cdot \vec{q}\,)} {m_{PS}^2 + q^2} \,(\vec{\tau}_i \cdot \vec{\tau}_j) \,, \label{PS} \\ 
V^{S}_{ij} (\vec{q}\,) &= -\frac{g_{ch}^2}{(2\pi)^3} \,\frac{\Lambda^2_{\chi SB}}{\Lambda^2_{\chi SB}+q^2} \,\frac{1}{m_S^2 + q^2} \,,
\end{align}
where $\vec{q}$ is the transferred three-momentum, the $\sigma$'s
($\tau$'s) are the spin (isospin) Pauli matrices whereas $m_q$, $m_{PS}$ and $m_S$ are, respectively, the masses of the quark, pseudoscalar and scalar bosons.

Above the chiral symmetry breaking scale there are no Goldstone boson interactions between quarks. However, quarks still interact perturbatively through multi-gluon exchanges described by the Lagrangian
\begin{equation}
\label{Lg}
{\mathcal L}_{gqq}= i \sqrt{4\pi \alpha _{s} } \,\,\overline{\psi } \gamma _{\mu} G^{\mu}_c \lambda _{c}\psi  \, ,
\end{equation}
where $\lambda _{c}$ are the SU(3) color matrices and $G^{\mu }_c$ the gluon field. The static approximation of the one-gluon exchange (OGE) interaction produces the following potential in momentum space:
\begin{align}
V_{ij}^{\text{OGE}}(\vec{q}\,) &= \frac{1}{(2\pi)^3} \frac{1}{4} (\vec{\lambda}_i^c\cdot\vec{\lambda}_j^c) 4\pi\alpha_s \Big\{ \frac{1}{q^2} \nonumber \\
&
- \frac{1}{4m_im_j} \big( 1+\frac{2}{3}(\vec{\sigma}_i\cdot \vec{\sigma}_j) \big) \nonumber \\
&
+ \frac{1}{4m_im_j} \frac{1}{q^2} [\vec{q}\otimes\vec{q}\,]^2\cdot[\vec{\sigma}_i\otimes\vec{\sigma}_j]^2 \Big\} \,.
\end{align}
The wide energy range needed to provide a consistent description of light, strange and heavy mesons requires an effective scale-dependent strong coupling constant. We use the frozen coupling constant~\cite{Vijande:2004he}:
\begin{equation}
\alpha_{s}(\mu_{ij})=\frac{\alpha_{0}}{\ln\left( 
\frac{\mu_{ij}^{2}+\mu_{0}^{2}}{\Lambda_{0}^{2}} \right)} \,,
\end{equation}
in which $\alpha_{0}$, $\mu_{0}$ and $\Lambda_{0}$ are parameters of the model determined by a global fit to the meson spectra~\cite{Segovia:2008zza, Segovia:2008zz, Segovia:2011dg, Segovia:2016xqb, Ortega:2020uvc}.

Note herein that, for the $N\overline N$ system, besides the direct diagrams, one must have a contribution from annihilation interaction. The real part of this can be derived from the one-gluon and one-pion exchanges, as explained in Ref.~\cite{Entem:2006dt}.

In order to extract the interaction at nucleon level, we use the Resonating Group Method (RGM)~\cite{Wheeler:1937zza, Tang:1978zz}. In this approach, the $N\overline N$ interaction appears as a residual effect of the underlying quark dynamics (see, for instance, Refs.~\cite{Fernandez:2019ses, Ortega2020, Martin-Higueras:2024qaw, Conde-Correa:2024qzh} for a detailed explanation).

The total wave function of the $N$ is written as
\begin{align}\label{fondaB}
\psi_N = \phi_N (\vec{p}_{\xi_1},\vec{p}_{\xi_2})\chi_N\xi_c[1^3]\,,
\end{align}
where $\xi_c[1^3]$ is the color singlet wave function, $\chi_N$ is the totally symmetric spin-isospin wave function, coupled to the quantum numbers of the nucleon, and $\phi_N(\vec{p}_{\xi_1},\vec{p}_{\xi_2})$ is the internal spatial baryon wave function, which is modeled as
\begin{align}\label{eq:internal}
\phi_N (\vec{p}_{\xi_1},\vec{p}_{\xi_2}) =
\left[\frac{2b^2}{\pi}\right]^{\frac{3}{4}}
e^{-b^2 p_{\xi_1}^2}
\left[\frac{3b^2}{2\pi}\right]^{\frac{3}{4}}
e^{-\frac{3b^2}{4} p_{\xi_2}^{2}} \,,
\end{align}
where $\vec p_{\xi_i}$ are the Jacobi coordinates and $b=0.518\,\text{fm}$ encodes the strong-size of the nucleon. This parameter was fixed in Ref.~\cite{Valcarce:1995dm} from an analysis of the three-quark system using the previous quark-quark potentials in the Born-Oppenheimer approach.

From RGM, integrating out all the internal degrees of freedom, the projected Schr\"odinger equation for the $N\overline N$ relative wave function can be written as
\begin{align}
\label{Schr}
\left( \frac{\left.\vec{P}^{'}\right.^{2}}{2\mu} - E \right) \chi(\vec{P}^{'})
+ \int{}^{\rm RGM} V_D(\vec{P}^{'}\!\!,\vec{P}_i) \chi(\vec{P}_i) d\vec{P}_i = 0 \,,
\end{align}
where $E= E_T-E_A-E_B$ is the relative energy of the clusters, and $^{\rm RGM}V_D (\vec{P}^{'}\!\!,\vec{P}_i)$ is the direct potential kernel, given by
\begin{align}
\label{eq:fullpot}
^{\rm RGM}\!V_D (\vec{P}^{'}\!\!,\vec{P}_i) =& \sum_{i \in A,j\in B} \int \phi_A^{*} (\vec{p}_{\xi_A^{'}}) \phi_B^{*} (\vec{p}_{\xi_B^{'}}) V_{ij}(\vec{P}^{'}\!\!,\vec{P}_i) \nonumber \\
&
\times \phi_A (\vec{p}_{\xi_A}) \phi_B (\vec{p}_{\xi_B}) d\vec{p}_{\xi_A^{'}} d\vec{p}_{\xi_B^{'}} d\vec{p}_{\xi_A} d\vec{p}_{\xi_B} \,,
\end{align}
where $\vec p_{\xi_{A(B)}^{(\prime)}}$ encodes the needed Jacobi momentum to define the internal wave function of the $N$ in Eq.~\eqref{eq:internal}, and $V_{ij}$ is the CQM potential between the quark $i$ and $j$ in the cluster $A$ and $B$, respectively.

The $N\overline N$ annihilation into mesons are processes whose description is intricate from a microscopic approach. Therefore, it is more convenient to parametrize them with a complex-like interaction that takes into account the loss of $N\overline N$ flux. Following Ref.~\cite{Entem:2006dt}, we include in our calculation an optical potential which does not have spin neither isospin dependence in the parameters, but gives a reasonable description of the $N\overline N\to N\overline N$ low-energy cross section. Its simplest expression is given by
\begin{align}
\label{eq:anhV}
V_{q\bar q}^{\rm Anh}(\vec q\,) = i\,\kappa\,e^{-\varsigma^{2}q^2} \,,
\end{align}
where the values of $\kappa$ and $\varsigma$ are taken from Ref.~\cite{Entem:2006dt}.

The poles emerging from the coupled-channels calculation will be obtained as zeros of the inverse of the $T$-matrix, calculated with the Lippmann-Schwinger equation
\begin{align} 
\label{ec:Tonshell}
T_\beta^{\beta'}(z;p',p) &= V_\beta^{\beta'}(p',p)+\sum_{\beta''}\int dq\,q^2\,
V_{\beta''}^{\beta'}(p',q) \nonumber \\
&
\times \frac{1}{z-E_{\beta''}(q)}T_{\beta}^{\beta''}(z;q,p) \,,
\end{align}
where $\beta$ represents the required set of quantum numbers to determine a partial wave in the $N\overline N$ channel, $V_{\beta}^{\beta'}(p',p)$ are the full RGM potentials from Eq.~\eqref{eq:fullpot} and $E_{\beta''}(q)$ is the non-relativistic energy for the momentum $q$. This equation is solved using the matrix-inversion method described in Ref.~\cite{Machleidt:1003bo}.


\begin{table}[!t]
\caption{\label{tab:model} Quark model parameters.}
\begin{ruledtabular}
\begin{tabular}{llr}
Quark masses & $m_{q}$ (MeV) & $313$ \\[2ex]

Goldstone Bosons & $m_{PS}$ $(\mbox{fm}^{-1})$ & $0.70$ \\
		  & $m_{S}$ $(\mbox{fm}^{-1})$ & $3.42$ \\
 		  & $\Lambda_{\chi SB}$ $(\mbox{fm}^{-1})$ & $4.20$ \\
 		  & $g^{2}_{ch}/4\pi$ & $0.54$ \\[2ex]

OGE & $\alpha_{0}$ & $2.118$ \\
    & $\Lambda_{0}$ $(\mbox{fm}^{-1})$ & $0.113$ \\
    & $\mu_{0}$ (MeV) & $36.976$ \\[2ex]


Optical potential & $\kappa$ (GeV$^{-2}$) & $-0.74$ \\
                  & $\varsigma$ (fm) & $0.49$ \\
\end{tabular}
\end{ruledtabular}
\end{table}

\section{RESULTS}
\label{sec:results}

A detailed discussion of the peculiarities of our coupled-channels calculation will be given in the following; however, a few initial comments are in order here. First, all discussed model parameters are summarized in Table~\ref{tab:model}. Second, there are two types of theoretical uncertainties in our results: one is intrinsic to the numerical algorithm and the other is related to the way the model parameters are fixed. The numerical error is negligible and, as mentioned above, the model parameters are adjusted to reproduce a certain number of hadron observables within a determinate range of agreement with experiment. It is therefore difficult to assign an error to these parameters and consequently to the quantities calculated using them. Finally, the experimental resonance parameters are obtained by a Breit-Wigner parametrization and one should be cautious when comparing them with pole positions.

We perform an analysis of the possible singularities in the $T$-matrix, and in the different Riemann sheets, that are located close to the $N\bar N$ threshold. We evaluate the sectors with total spin $J=\{0,1,2\}$, in different partial waves: $^1S_0$, $^3S_1-{}^3D_1$, $^1P_1$, $^3P_0$, $^3P_1$ and $^3P_2-{}^3F_2$. Since the $p\bar p$ and $n\bar n$ thresholds have a sizable separation, $m_{n\bar n}-m_{p\bar p}=(1879.13-1876.54)\,\text{MeV}=2.59\,\text{MeV}$, the calculation can be sensitive to possible isospin breaking effects and thus we work in charged basis, taking the combinations
\begin{subequations}
\begin{align}
|p\bar p\rangle &= \frac{1}{\sqrt{2}} (|N\bar N\rangle_1 + |N\bar N\rangle_0) \,,\\
|n\bar n\rangle &= \frac{1}{\sqrt{2}} (|N\bar N\rangle_1 - |N\bar N\rangle_0) \,.
\end{align}
\end{subequations}
That is to say, the potentials are defined as
\begin{subequations}
\begin{align}
V_{p\bar p}=V_{n\bar n} = \frac{1}{2}(V_1+V_0) \,,\\
V_{p\bar p\to n\bar n}=V_{n\bar n \to p\bar p} = \frac{1}{2}(V_1-V_0) \,.
\end{align}
\end{subequations}

\begin{table*}[t!]
\caption{\label{tab:states} The $N\bar N$ poles emerging from the coupled-channels calculation without taking into account the optical potential, $V_{q\bar q}^{\rm Anh}$. \emph{1$^{st}$-2$^{nd}$ columns:} Quantum numbers of the sector; \emph{3$^{rd}$ column: } Mass of the state; \emph{4$^{th}$ column: } Width of the state; \emph{5$^{th}$ column:} Binding energy with respect to the lower threshold (\emph{i.e.} $p\bar p$-threshold); \emph{6$^{th}$ column:} Riemann sheet of each $(N\bar N)$-channel ($p\bar p$, $n\bar n$), where $F$ denotes first and $S$ second Riemann sheets; \emph{7$^{th}$-8$^{th}$ columns:} Probability of the charged channels; \emph{9$^{th}$-10$^{th}$ columns:} Probability of the isospin channels. Note that the probabilities of each channel are shown only for either bound-states or resonances.}
\begin{ruledtabular}
\begin{tabular}{cccccc|cc|cc}
 $J^{PC}$ & $^{2S+1}L_J$ & Mass & Width & $E_B$ & RS & ${\cal P}_{p\bar p}$ & ${\cal P}_{n\bar n}$ & ${\cal P}_{I=0}$ & ${\cal P}_{I=1}$ \\
 \hline
 $0^{-+}$ & $^1S_0$ & $1873.0$ & $0.0$ & $-3.5$ & (S,F) &  &  &  &  \\
 $0^{++}$ & $^3P_0$ & $1876.4$ & $0.0$ & $-0.1$ & (F,F) & $60.5$ & $39.5$ & $96.7$ & $3.3$ \\
 $1^{--}$ & $^3S_1-{}^3D_1$ & $1876.1$ & $0.0$ & $-0.4$ & (F,F) & $81.1$ & $18.9$ & $84.3$ & $15.7$ \\
 $1^{+-}$ & $^1P_1$ & $1876.2$ & $148.3$ & $-0.3$ & (S,S) & & & & \\
 $1^{++}$ & $^3P_1$ & $1864.2$ & $80.7$ & $-12.3$ & (S,S)& & & & \\
 $2^{++}$ & $^3P_2-{}^3F_2$ & $1886.3$ & $58.8$ & $+9.7$ & (S,S) & $51.7$ & $48.3$ & $99.9$ & $0.1$ \\
\end{tabular}
\end{ruledtabular}
\end{table*}

If we first analyze the $N\bar N$ system without the optical potential defined in Eq.~\eqref{eq:anhV}, our results are shown in Table~\ref{tab:states}. We only find two bound states, one in the $J^{PC}=1^{--}$ sector with partial waves $^3S_1-{}^3D_1$ whereas the other with quantum numbers $J^{PC}=0^{++}$ and partial wave $^3P_0$. Both are originally isoscalar bound states which acquire an isovector component because the experimental mass difference between the $p\bar p$ and $n\bar n$ thresholds. Their binding energy (defined as $E_B=E-m_{p\bar p}$) is quite small: about $-0.4\,\text{MeV}$ for the $J^{PC}=1^{--}$ case and $-0.1\,\text{MeV}$ for the $0^{++}$ one. Besides these two bound states, we find an isoscalar resonance in the $J^{PC}=2^{++}$ and three virtual states in the $J^{PC}=0^{-+}$, $1^{+-}$ and $1^{++}$ sectors.

These poles could be responsible of the recently announced plethora structures around the $p\bar p$ threshold: $X(1835)$, $X(1840)$, $X(1870)$, $X(1880)$, $\ldots$ However, prior to a rush assignment, we must take into account the optical potential which deals with the imaginary part of the $N\overline N$ annihilation contribution. When we put it into operation, the poles evolve from their original positions shown in Table~\ref{tab:states} to those reported in Table~\ref{tab:states2}.

\begin{table}[t!]
\centering
\caption{\label{tab:states2} The $N\bar N$ poles emerging from the coupled-channels calculation taking into account the optical potential, $V_{q\bar q}^{\rm Anh}$. \emph{1$^{st}$-2$^{nd}$ columns:} Quantum numbers of the sector; \emph{3$^{rd}$ column: } Mass of the state; \emph{4$^{th}$ column: } Width of the state; \emph{5$^{th}$ column:} Binding energy with respect to the lower threshold (\emph{i.e.} $p\bar p$-threshold); \emph{6$^{th}$ column:} Riemann sheet of each $(N\bar N)$-channel ($p\bar p$, $n\bar n$), where $F$ denotes first and $S$ second Riemann sheets.}
\begin{ruledtabular}
\begin{tabular}{cccccc}
 $J^{PC}$ & $^{2S+1}L_J$ & Mass & Width & $E_B$ & RS \\
 \hline
 $0^{++}$ & $^3P_0$ & $1896.2$ & $33.9$ & $19.6$ & (F,F)  \\
          &         & $1873.1$ & $13.1$ & $-3.5$ & (S,F)  \\
          &         & $1876.5$ & $8.6$ & $-0.1$ & (S,S)  \\
 $1^{--}$ & $^3S_1-{}^3D_1$ & $1896.2$ & $8.1$ & $19.7$ & (F,F) \\
          &                 & $1896.4$ & $7.6$ & $19.8$ & (S,F)  \\
          &                 & $1889.6$ & $128.6$ & $13.0$ & (S,S)  \\
 $1^{+-}$ & $^1P_1$ & $1870.6$ & $13.9$ & $-6.0$ & (S,S)  \\
          &         & $1861.3$ & $39.1$ & $-15.3$ & (S,S) \\
 $1^{++}$ & $^3P_1$ & $1869.4$ & $20.4$ & $-7.1$ & (S,S)\\
 $2^{++}$ & $^3P_2-{}^3F_2$ & $1870.6$ & $30.9$ & $-5.9$ & (S,S) \\
\end{tabular}
\end{ruledtabular}
\end{table}

Among the results shown in Table~\ref{tab:states2}, two are of particular interest. On one hand, the number of poles is larger compared to the zero-optical-potential situation and, on the other hand, all poles move to non-physical Riemann sheets, \emph{i.e.} there are no bound states. 

In more detail, the $J^{PC}=0^{-+}$ virtual state is washed out, moving far away from the $p\bar p$ threshold. The $1^{++}$ virtual pole moves closer to the $p\bar p$ threshold whereas the $1^{+-}$ virtual state reduces its energy to $\sim 1.86\,\text{GeV}$, while a new one with smaller width appears. In the $2^{++}$ sector, the resonance dives under the $n\bar n$ and $p\bar p$ channels, and it becomes a virtual state located at about $6\,\text{MeV}$ below the $p\bar p$ threshold. 

\begin{figure}[t!]
\includegraphics[width=.5\textwidth]{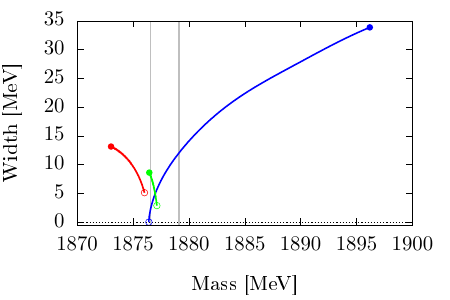}
\caption{\label{pole3P0} Pole trajectory of $J^{PC}=0^{++}$ poles in the Riemann sheet (F,F) (blue), (S,S) (green) and (S,F) (red) with increasing values of the optical potential $\kappa$ parameter. The vertical gray lines represent the $p\bar p$ and $n\bar n$ thresholds. The opened circles show the initial position at $\kappa=0$ GeV$^{-2}$ for (F,F), $-0.14$ GeV$^{-2}$ for (S,S) and $-0.17$ GeV$^{-2}$ for (S,F) pole. The closed circles indicate the final value at $\kappa=-0.74$ GeV$^{-2}$.}
\end{figure}

\begin{figure}[t!]
\includegraphics[width=.5\textwidth]{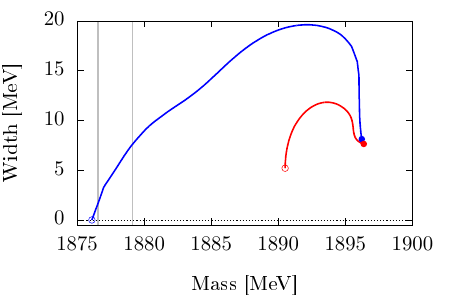}
\caption{\label{pole3S1} Pole trajectory of $J^{PC}=1^{--}$ poles in the Riemann sheet (F,F) (blue) and (S,F) (red) with increasing values of $\kappa$. The vertical gray lines represent the $p\bar p$ and $n\bar n$ thresholds.  The opened circles show the initial position at $\kappa=0$ GeV$^{-2}$ for (F,F) and $-0.18$ GeV$^{-2}$ for (S,F) pole. The closed circles indicate the final value at $\kappa=-0.74$ GeV$^{-2}$.}
\end{figure}

Let us now focus on the $J^{PC}=0^{++}$ and $1^{--}$ sectors, where bound states were detected when the optical potential was ignored. These bound states acquire widths and higher masses, moving above the $p\bar p$ and $n\bar n$ thresholds, still in the first Riemann sheet of each $N\overline N$-channel ($p\bar p$, $n\bar n$). The $0^{++}$ singularity ends up with a mass of $1896\,\text{MeV}$ and a width of $33.9\,\text{MeV}$, while the $1^{--}$ is located at a mass of $1896\,\text{MeV}$ but it has a much smaller width of $8.1\,\text{MeV}$. Their trajectory in the complex energy plane can be found in Figs.~\ref{pole3P0} and~\ref{pole3S1} for $0^{++}$ and $1^{--}$, respectively. As they move to nonphysical regions, they cannot be seen as Breit-Wigner structures, only their tails could be seen below the $p\bar p$ threshold.

Besides the original $0^{++}$ and $1^{--}$ poles appearing in the first Riemann sheet of each $N\overline N$-channel ($p\bar p$, $n\bar n$), two additional poles appear in the (S,F) and (S,S) Riemann sheets for both sectors.\footnote{$F$ denotes first and $S$ second Riemann sheets.} For the $0^{++}$ sector, these two poles are below the $p\bar p$ threshold, so they act as virtual states whose influence could be seen above the $p\bar p$ and $n\bar n$ thresholds. In the case of the $1^{--}$ sector, the (S,F) pole is close to the (F,F) one, at around $20\,\text{MeV}$ above the $p\bar p$ threshold. On the contrary, the (S,S) pole is a resonance with a large width of $\sim 130\,\text{MeV}$.

\section{Summary}
\label{sec:summary}

In this work we have revisited and updated the perturbative calculation shown in Ref.~\cite{Entem:2006dt} where an analysis of the possible existence of bound states or near-threshold resonances in proton-antiproton scattering, as well as energy shifts in protonium, was performed. 

The $N\overline{N}$ potential used here is derived from a $G$-parity transformation of the quark-model-based $NN$ interaction presented in~\cite{Entem:2000mq, Valcarce:2005em}. The additional $N\overline{N}$ annihilation contribution follows the approach in Ref.~\cite{Entem:2006dt}, where it is described by a complex phenomenological potential. The real part arises from one-pion and one-gluon exchange annihilations, while the imaginary part is modeled as an energy-independent Gaussian potential.

If the optical potential is disregarded, two shallow bound states are found in the $J^{PC}=0^{++}$ and $1^{--}$ sectors. Both are predominantly isoscalars whose isovector component is acquired because the experimental mass difference between the $p\bar p$ and $n\bar n$ thresholds. Besides, a resonance in the $J^{PC}=2^{++}$ and three virtual states in the $J^{PC}=0^{-+}$, $1^{+-}$ and $1^{++}$ sectors are also found. When the optical potential is included, the poles evolve from their original positions in such a way that there are no bound states and a larger number of $T$-matrix poles appear.

It is still premature to make assignments between our theoretical states and the signals observed experimentally. Our model, however, predicts poles above (below) the $N\overline N$ threshold within non-physical regions, whose effects manifest as enhancements below (above) the aforementioned threshold. This may lead to a sudden shift in the $p\bar p$ line shape, similar to what was observed in Ref.~\cite{BESIII:2016fbr}. These findings suggest that nucleon-antinucleon interactions are complex, with a wealth of scattering singularities near the proton-antiproton and neutron-antineutron thresholds.


\begin{acknowledgments}
Work partially financed by 
EU Horizon 2020 research and innovation program, STRONG-2020 project, under grant agreement No. 824093;
Ministerio Espa\~nol de Ciencia e Innovaci\'on under grant Nos. PID2019-105439GB-C22 and PID2022-140440NB-C22;
Junta de Castilla y León under grant No. SA091P24;
Junta de Andaluc\'ia under contract Nos. PAIDI FQM-370 and PCI+D+i under the title: ``Tecnolog\'\i as avanzadas para la exploraci\'on del universo y sus componentes" (Code AST22-0001).
\end{acknowledgments}


\bibliography{draft-NNbar}

\end{document}